\documentclass[9pt,twocolumn,twoside]{osajnl}
\journal{ol}

\usepackage{comment}
\usepackage{txfonts}
\usepackage{amsmath,amssymb}
\usepackage{graphicx}
\setboolean{shortarticle}{true}
\usepackage[detect-weight]{siunitx}

\title{Ultrafast circularly polarized pulses tunable from the vacuum to deep ultraviolet}
\author[1*]{Athanasios Lekosiotis}
\author[1]{Christian Brahms}
\author[1]{Federico Belli}
\author[1]{Teodora F. Grigorova}
\author[1]{John C. Travers}

\affil[1]{School of Engineering and Physical Sciences, Heriot-Watt University, Edinburgh, EH14 4AS, UK}

\affil[*]{Corresponding author: a.lekosiotis@hw.ac.uk}

\begin{abstract}
We experimentally demonstrate the efficient generation of circularly polarized pulses tunable from the vacuum to deep ultraviolet (160-380~nm) through resonant dispersive wave emission from optical solitons in a gas-filled hollow capillary fiber. In the deep ultraviolet we measure up to \SI{13}{\micro\joule} of pulse energy, and from numerical simulations, we estimate the shortest output pulse duration to be 8.5~fs. We also experimentally verify that simply scaling the pulse energy by 3/2 between linearly and circularly polarized pumping closely reproduces the soliton and dispersive wave dynamics. Based on previous results with linearly polarized self-compression and resonant dispersive wave emission, we expect our technique to be extended to produce circularly polarized few-fs pulses further into the vacuum ultraviolet, and few to sub-fs circularly polarized pulses in the near-infrared.
\end{abstract}

\setboolean{displaycopyright}{False}

\begin{document}

\maketitle

Ultrafast ultraviolet (UV) laser pulses with circular polarization are an important tool in circular dichroism spectroscopy \cite{comby_real-time_2018}, attosecond science \cite{guo_spatiotemporal_2019}, and medicine \cite{kunnen_application_2015}. The generation of tunable deep UV (DUV, 200-300~nm) and vacuum UV (VUV, $<200~$nm) pulses with linear polarization has been widely reported \cite{baum_generation_2004, travers_high-energy_2019, belli_broadband_2020}. The generation of circularly polarized pulses in the UV is more challenging, and is usually obtained via a two-step process requiring the use of a frequency conversion stage to generate UV with linear polarization followed by a second stage for polarization conversion \cite{graf_intense_2008}. The maximum bandwidth, and hence minimum pulse duration, which can be achieved in this way is limited by both the up-conversion process and the lack of achromatic phase retardation in the ultraviolet. Recently, we demonstrated direct frequency up-conversion of circularly polarized laser pulses via four-wave mixing in hollow capillary fibers (HCF) \cite{lekosiotis_generation_2020}. We generated high-energy ($27~\muup$J) circularly polarized pulses at 266~nm, avoiding the need for UV waveplates. While this approach provides tunability through nonlinear phase modulation or by seeding with a tunable infrared signal, it requires dual-color input, and to reach the VUV, a DUV pump pulse is needed.

An alternative approach to ultrafast UV pulse generation is resonant dispersive wave (RDW) emission in gas-filled HCF ~\cite{travers_high-energy_2019}. This is a single-stage, single-color (near-infrared) pump scheme with the ability to directly generate few-fs, tunable VUV and DUV pulses with high efficiency (over 15\%). Wavelength tuning is achieved simply by altering the gas pressure. In this letter we demonstrate that this technique also works for circular polarization with minimal modification by generating ultrafast circularly polarized pulses from the VUV and across the DUV (160-380~nm) with high energy (up to \SI{13}{\micro\joule}). This approach can be extended further into the VUV~\cite{travers_high-energy_2019}, and to the generation of few to sub-fs circularly polarized pulses in the infrared \cite{brahms_infrared_2020}.

\begin{figure*}[th]
\centering
\includegraphics[width=1\linewidth]{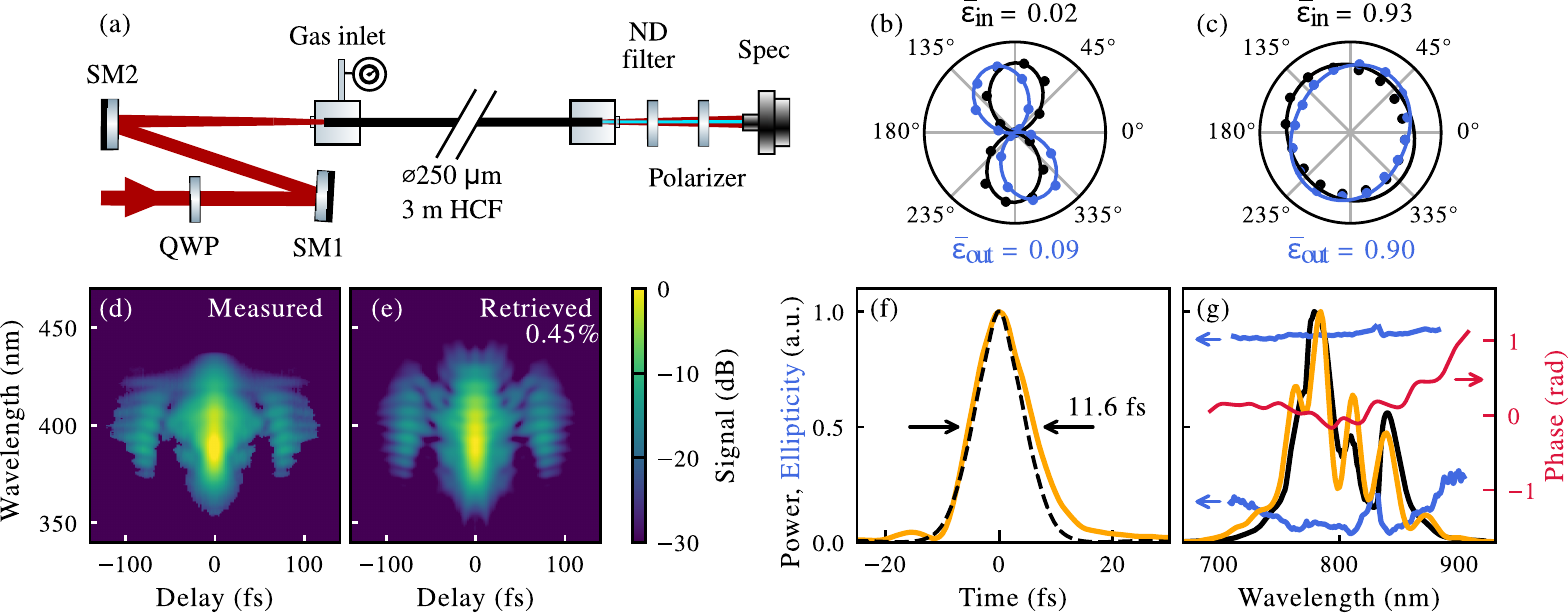}
\caption[Input pulse]%
{(a) Sketch of the experimental apparatus. A quarter-wave plate (QWP) converts the 790~nm pulses to circular polarization. The beam is focused with a concave mirror (SM2) to the input facet of a 3~m long, $250~\muup$m core diameter hollow capillary fiber (HCF). A neutral density (ND) filter, a Rochon polarizer and a CCD spectrometer are used for polarization measurements of the output pulses. Angle-dependent polarizer transmission for the linearly polarized (b) and circularly polarized (c) signal at 790~nm, measured before SM1 (black dots) and at the output of the evacuated HCF (blue dots). Solid lines show the fitting curves and $\bar{\varepsilon}_{\text{in}}$, $\bar{\varepsilon}_{\text{out}}$ are the corresponding averaged ellipticities. (d, e) Measured and reconstructed SHG-FROG traces of the input pulses on a logarithmic color scale. (f, g) Retrieved pulse and spectral profile (11.6~fs, orange), Fourier transform-limited pulse (9.7~fs, dashed black), externally measured power spectrum (solid black) and retrieved spectral phase (red). The circular (top blue) and linear (bottom blue) polarization profiles are measured at the output of the evacuated HCF.}
\label{fig:fig1}
\end{figure*}

The experimental setup is identical to that used for RDW generation with linearly polarized pulses~\cite{travers_high-energy_2019}, except for the addition of a quarter-wave plate (QWP) for conversion to circular polarization, as shown in Fig.~\ref{fig:fig1}(a). In brief, the 30~fs FWHM (full width at half maximum) 790~nm output of a Ti:Sapph laser source operating at 1~kHz repetition rate is compressed to 11.6~fs via spectral broadening in a helium-filled stretched HCF followed by spectral phase compensation with chirped mirrors and silica wedges. We use our own design for HCF stretching to obtain long HCF lengths with negligible bend loss~\cite{nagy_flexible_2008}. A motorized half-wave plate (HWP) and a silicon plate are used for conversion to vertical linear polarization and power control. The circularly polarized pulses exiting the QWP are focused with a concave mirror (SM2) at the input facet of a 3~meter-long, $250~\muup$m core diameter HCF, which is mounted in a gas cell sealed with 1~mm thick magnesium fluoride (MgF$_2$) windows at each end. At the fiber output, a broadband linear Rochon polarizer is used for polarization measurements with a power attenuator (neutral-density (ND) filter) to avoid optical damage. Both polarization and spectral energy density (SED) measurements are recorded with a system consisting of a commercial fiber-coupled charge-coupled device (CCD) spectrometer equipped with an integrating sphere. This system is calibrated as a whole using NIST-traceable lamps, and the SED calibration is verified by comparing with direct energy measurements. For spectrum measurements in the VUV, the windowless HCF output is connected to a vacuum system containing a home-built VUV spectrometer \cite{travers_high-energy_2019}.

Our method for characterizing the polarization is based on measuring and analyzing the spectrum transmitted through the polarizer as a function of its rotation angle \cite{lekosiotis_generation_2020}. Fitting the recorded SED to the generalized Malus' law \cite{azzam_ellipsometry_1987} yields the wavelength-dependent ellipticity $\varepsilon (\lambda)$, which ranges from 0.0 for linear polarization to 1.0 for pure circular polarization. This is subsequently weighted with respect to the power spectrum to obtain the spectrally averaged ellipticity $\bar{\varepsilon}$. Polarization measurements of the pump pulses are shown in Fig.~\ref{fig:fig1}(b, c) for data recorded without the QWP (linearly polarized) and with the QWP (circularly polarized). In both cases, the ellipticity is measured at two positions; once before the steering mirrors but after the QWP, when present ($\bar{\varepsilon}_{\text{in}}$), and once at the output of the evacuated HCF ($\bar{\varepsilon}_{\text{out}}$). From the former, we obtain the polarization induced by the variable attenuator (HWP and silicon plate) and the QWP (for circular polarization), and from the latter we obtain the closest estimate to the polarization of the pulses coupled into the HCF, as transmission through the windows and the stretched HCF induces negligible birefringence. 

The two polar plots in Fig.~\ref{fig:fig1}(b, c) show the signal at the central wavelength (790~nm) as a function of the polarizer rotation angle at both measurement positions. Both the linearly and the circularly polarized systems exhibit a slight polarization change from their $\bar{\varepsilon}_{\text{in}}$ measurements when measured at the output of the evacuated HCF. For linearly polarized light, the averaged ellipticity increases from 0.02 to 0.09 and the polarization ellipse rotates in the anti-clockwise direction, while for circularly polarized light, the averaged ellipticity slightly decreases from 0.93 to 0.90 and the polarization ellipse slightly shifts in the clockwise direction. The ellipticity change between the two measurement positions is mainly induced by the oblique incidence of light on the two metal steering mirrors. In Fig.~\ref{fig:fig1}(g), we show the ellipticity measured at the output of the evacuated HCF for circularly polarized (top blue) and linearly polarized (bottom blue) pulses and observe spectrally uniform polarization profiles in both cases.

\begin{figure*}[th]
\centering
\includegraphics[width=1\linewidth]{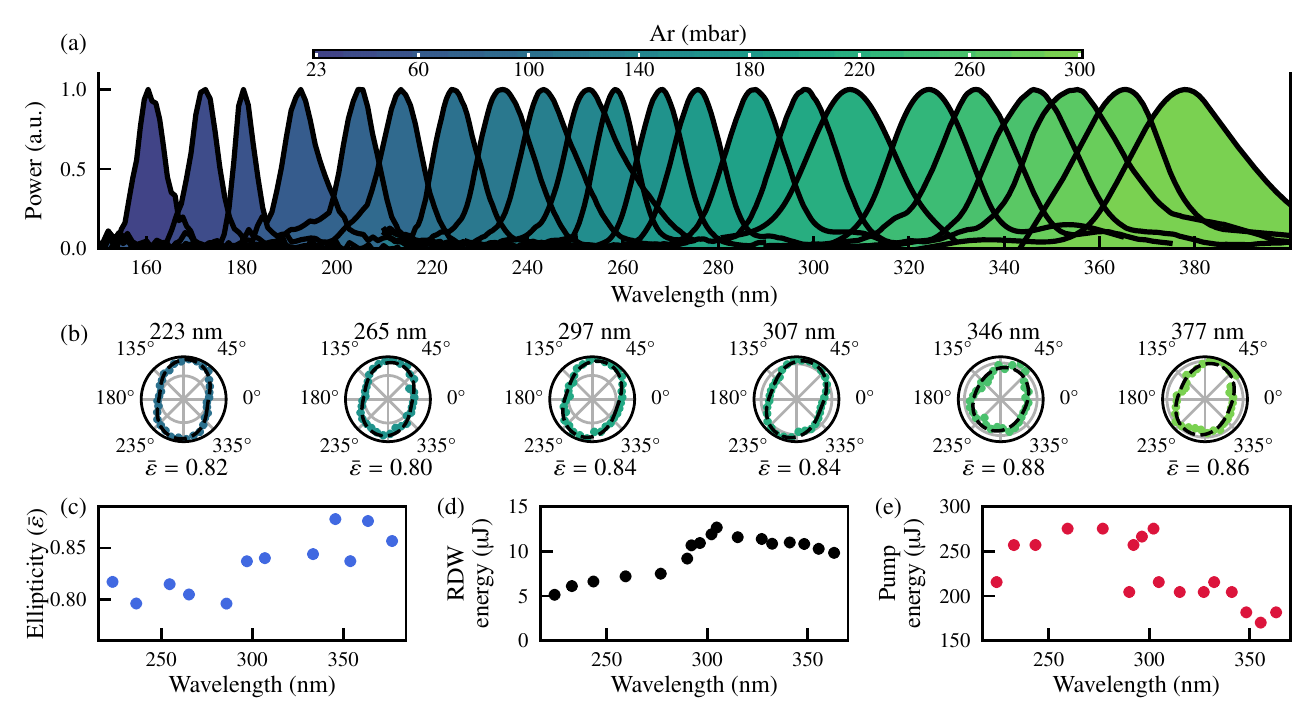}
\caption[RDWs]%
{(a) Experimental RDW spectra obtained by tuning the pressure inside the HCF between 23~mbar (shortest wavelength) and 300~mbar (longest wavelength) of argon. (b) Angle-dependent polarizer transmission of the RDW signal at the central wavelength for a selection of generation conditions. $\bar{\varepsilon}$ is the spectrally averaged ellipticity and dashed lines show fits to the data. (c) RDW averaged ellipticity, (d) output RDW energy and (e) corresponding coupled pump energy as a function of the DUV emission wavelengths.} 
\label{fig:fig2}
\end{figure*}

We perform temporal characterization of the compressed 790~nm pulses via second-harmonic generation frequency-resolved optical gating (SHG-FROG), as shown in Fig.~\ref{fig:fig1}(d-g). The pulse duration at the HCF entrance (including transmission through the input window and the QWP) is retrieved as 11.6~fs FWHM duration with a 9.7~fs Fourier transform limit (FTL). The retrieval error is 0.45\% on a $256 \times 256$ time-frequency grid. In Fig.~\ref{fig:fig1}(g), the FROG-retrieved spectrum is compared with an external spectrum measurement to verify the retrieval. 

We fill the HCF with between 23 and 300 mbar of argon (Ar) and pump the system with the compressed circularly polarized pulses (0.90 ellipticity). For each pressure we optimize the RDW emission by adjusting the pump pulse energy. The dispersive-wave spectra are shown in Fig.~\ref{fig:fig2}(a). The circularly polarized pulses can be tuned continuously between 160~nm and 380~nm, with increasing pressure leading to longer emission wavelengths, as previously observed with linear polarization~\cite{travers_high-energy_2019}. In the polarization measurements of the generated RDWs in the DUV, we notice a polarization rotation in the clockwise direction [see Fig.~\ref{fig:fig2}(b)] and a slight increase of the averaged ellipticity from 0.80 to 0.88 [see Fig.~\ref{fig:fig2}(c)] for increasing gas pressure. The rotation of the ellipse is consistent with that of the circularly polarized pump pulse measured at the output of the evacuated HCF stage [see Fig.~\ref{fig:fig1}(c), blue line], especially at low pressures. This is evidence of the underlying polarization transfer from the pump to the RDW pulses.

In Fig.~\ref{fig:fig2}(d, e) we show the output energy of the circularly polarized DUV RDW pulses and the coupled pump energy used to maximize the RDW energy at each pressure, while maintaining a smooth Gaussian-like spectral profile. The RDW energy at shorter DUV wavelengths is around $5~\muup$J, increasing to $13~\muup$J at around 300 nm. The conversion efficiency (defined as the energy ratio of the emitted RDW to the coupled pump) is 2\% at shorter and 6\% at longer emission wavelengths, as the optimal pump energy shifts from around 275~$\muup$J at shorter to 200~$\muup$J at longer wavelengths. By further increasing the pump energy, RDW emission down to 110 nm can be achieved~\cite{travers_high-energy_2019}.

\begin{figure}[ht]
\centering
\includegraphics[width=1\linewidth]{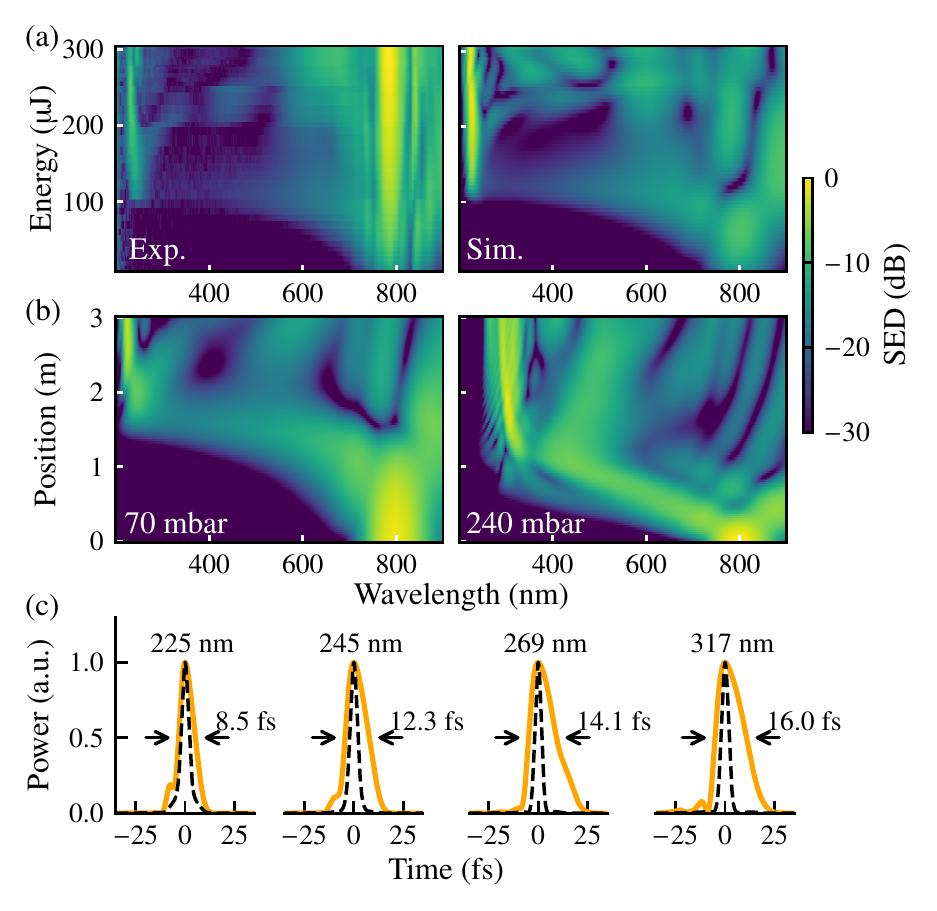}
\caption[Simulations]%
{(a) Experimental (left) and simulated (right) output spectra with pump energy for 70 mbar of argon. (b) Simulated evolution inside the HCF for 70 mbar (left) and 240 mbar (right) of argon. (c) Simulated RDW pulses at the HCF output with annotated central wavelengths and FWHM durations. Dashed lines show the transform-limited pulses (3.5~fs duration at all pressures).}
\label{fig:fig3}
\end{figure}

In Fig.~\ref{fig:fig3}(a), we compare experimental (left) and simulated (right) output spectra as the pump energy is changed for 70~mbar of argon. The pulse propagation model used for numerical simulations is discussed in ref.~\cite{travers_high-energy_2019} and includes a full vector polarization model. We simulate propagation in two orthogonally polarized degenerate fundamental modes, with initial conditions corresponding to a purely circularly polarized, transform-limited 11.6~fs Gaussian pulse centred at 790~nm. The energy coupled into the HCF is determined from the transmission of the evacuated HCF after accounting for linear propagation losses. The simulations reproduce the experimental measurements, especially the RDW emission dynamics (energy onset, spectral position and evolution). The difference in the pump spectral region is largely due to an amplified spontaneous emission pedestal originating from the laser system that does not contribute to the nonlinear dynamics.

Although we have not measured the ellipticity of the RDW pulses in the VUV, as this is not possible with our current apparatus, we expect that they are also circularly polarized, primarily because they are generated in the same way. Furthermore, the ellipticity measurements for the DUV RDW pulses are very consistent, with little wavelength dependence, and the agreement between experiment and simulations suggests that our model fully captures the relevant dynamics.

A direct experimental measurement of the temporal profile of circularly polarized UV pulses is also extremely challenging. However, we can infer the temporal dynamics using numerical simulations. The temporal profiles of the simulated circularly polarized RDW pulses are shown in Fig.~\ref{fig:fig3}(c), as obtained by band-pass filtering with 15\% relative bandwidth around the peak of the RDW signal. These pulses correspond to the pressure and energy of several of the experimental data in Fig.~\ref{fig:fig2}. The output pulse durations range between 8.5~fs and 16~fs FWHM between 225 and 317~nm, with a 3.5~fs FTL at all wavelengths. Shorter pulse durations are obtained when the RDW pulse is generated closer to the HCF exit and hence accumulate less dispersion, which occurs for lower emission wavelengths for our parameters. Fig.~\ref{fig:fig3}(b) confirms this, showing the spectral evolution inside the 3~m long HCF for 70~mbar (left) and 240~mbar (right) and pump energies of $250~\muup$J and $275~\muup$J, respectively. For 70~mbar, RDW emission occurs at 1.5~m, whereas for 240~mbar emission occurs at 0.7~m. By optimising the fiber length, the shortest pulses can be obtained at the output, however they would still need to pass through a dispersive window. Circularly polarized few-fs pulses, close to the FTL, could be delivered to vacuum for all RDW emission wavelengths by using a decreasing pressure gradient along the fiber length, as demonstrated for linear polarization~\cite{brahms_resonant_2020}.

\begin{figure}[ht]
\centering
\includegraphics[width=1\linewidth]{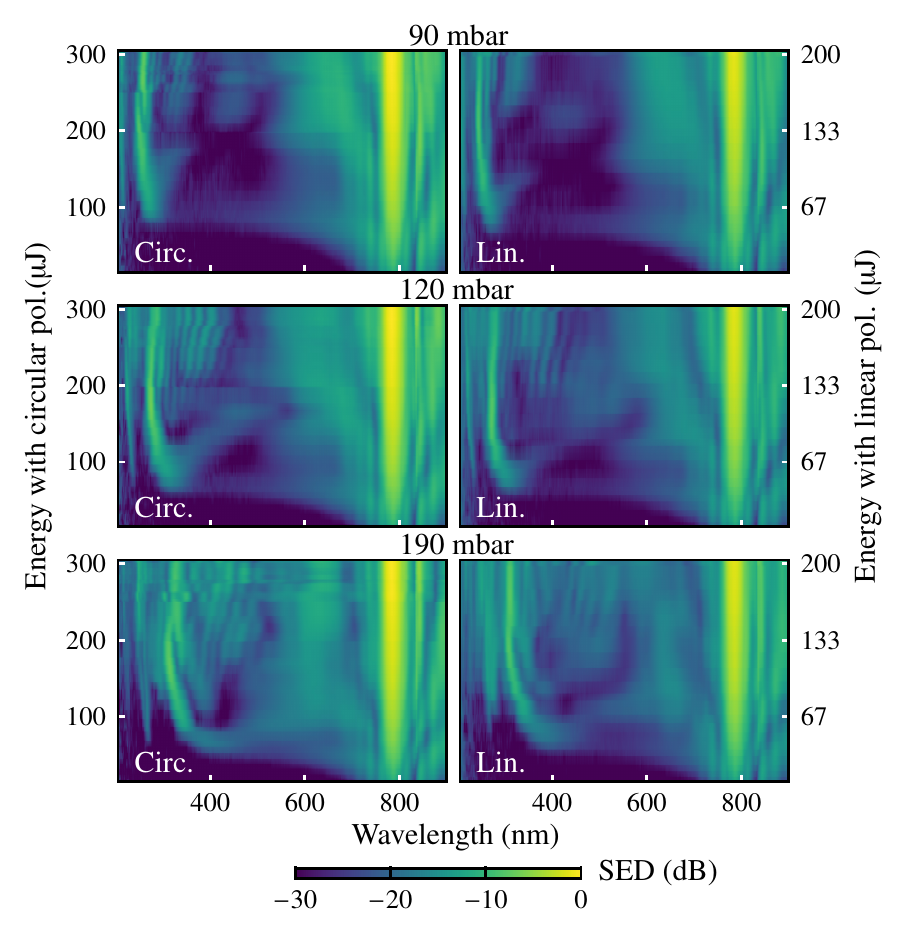}
\caption[Linear vs circular]%
{Experimental output spectra with circularly polarized (left) and linearly polarized (right) input pulses for 90 mbar (top), 120 mbar (middle) and 190 mbar (bottom) of argon. Pump energy is up to $300~\muup$J for circular polarization and up to $200~\muup$J for linear polarization to match the nonlinearity inside the HCF.} 
\label{fig:fig4}
\end{figure}

It is well known that the Kerr nonlinearity for circularly polarized pulses is $2/3$ that for linearly polarized pulses~\cite{boyd_nonlinear_2008,schimpf_circular_2009}. Since HCF is perfectly radially symmetric, its dispersion is independent of polarization state and hence this scaling transfers directly to soliton dynamics. As a result, one should be able to reproduce the linearly polarized output for a given gas pressure if the circularly polarized pump energy is increased by $3/2$. We make this comparison in Fig.~\ref{fig:fig4} for 90, 120 and 190 mbar of argon. At each pressure we show the experimental output spectra for circularly polarized input (left) obtained using the whole energy range available (up to $300~\muup$J) as well as the equivalent linearly polarized spectra (right) obtained using $2/3$ of the energy (up to $200~\muup$J). The spectral evolution in the two polarization regimes is almost identical for all pressures applied: we measure the same energy onset and evolution for the RDW pulse and very similar evolution for the blue-shifting pump, even after the formation of a supercontinuum at higher energies and pressures.

This verification that soliton dynamics demonstrated with linear polarization can be reproduced almost exactly with circular polarization at $3/2$ higher energy implies that all of the techniques we have demonstrated for linearly polarized soliton dynamics in HCF will also apply for circular polarization. This includes spectral extension of the RDW further into the VUV by tuning the gas pressure~\cite{travers_high-energy_2019} and to the visible and NIR by changing the driving wavelength~\cite{brahms_infrared_2020}, as well as the generation of sub-fs transients in the visible and infrared through soliton self-compression~\cite{travers_high-energy_2019, brahms_infrared_2020}. Furthermore, scaling of the energy and length required is feasible by adjusting the HCF core diameter~\cite{brahms_high-energy_2019} and direct delivery of compressed RDW pulses to vacuum can be achieved by using a gas pressure gradient~\cite{brahms_resonant_2020}. 

In summary, we have experimentally demonstrated the efficient generation of bright and tunable pulses in the ultraviolet with circular polarization via resonant dispersive wave emission in gas-filled HCF. We have also numerically estimated that the output RDW pulses have a short ($<16$~fs) duration. Furthermore, we have confirmed that routes to the generation of shorter pulses, and spectral extension across both the visible and near-infrared regions, are readily accessible using techniques demonstrated for linear polarization.

\section*{Acknowledgements}
This work was funded by the European Research Council (ERC) under the European Union's Horizon 2020 research and innovation program: Starting Grant agreement HISOL, No. 679649.

\section*{Disclosures}
The authors declare no conflicts of interest.


\bibliography{Paper_cHISOL}
\bibliographyfullrefs{Paper_cHISOL}

\end{document}